\documentclass[10pt,conference,twocolumn]{IEEEtran}

\usepackage{cite}
\usepackage[T1]{fontenc}
\usepackage{graphicx}
\usepackage{amssymb}
\usepackage{amsmath}
\usepackage{paralist}
\usepackage{setspace}
\usepackage{microtype}
\usepackage{balance}
\usepackage[caption=false,font=footnotesize]{subfig}
\usepackage{xcolor}
\usepackage{accents}
\usepackage{dsfont}
\usepackage{lipsum}
\usepackage{booktabs}
\usepackage[nolist,nohyperlinks]{acronym}
\usepackage{todonotes}
\usepackage{balance}

\newcommand{\snr}{\rho}

\usepackage{amssymb}
\usepackage{amsfonts}
\usepackage{mathrsfs}
\usepackage{xspace}
\usepackage{bm}
\usepackage{upgreek}

\newcommand{\safemath}[2]{\newcommand{#1}{\ensuremath{#2}\xspace}}

\safemath{\bma}{\mathbf{a}}
\safemath{\bmb}{\mathbf{b}}
\safemath{\bmc}{\mathbf{c}}
\safemath{\bmd}{\mathbf{d}}
\safemath{\bme}{\mathbf{e}}
\safemath{\bmf}{\mathbf{f}}
\safemath{\bmg}{\mathbf{g}}
\safemath{\bmh}{\mathbf{h}}
\safemath{\bmi}{\mathbf{i}}
\safemath{\bmj}{\mathbf{j}}
\safemath{\bmk}{\mathbf{k}}
\safemath{\bml}{\mathbf{l}}
\safemath{\bmm}{\mathbf{m}}
\safemath{\bmn}{\mathbf{n}}
\safemath{\bmo}{\mathbf{o}}
\safemath{\bmp}{\mathbf{p}}
\safemath{\bmq}{\mathbf{q}}
\safemath{\bmr}{\mathbf{r}}
\safemath{\bms}{\mathbf{s}}
\safemath{\bmt}{\mathbf{t}}
\safemath{\bmu}{\mathbf{u}}
\safemath{\bmv}{\mathbf{v}}
\safemath{\bmw}{\mathbf{w}}
\safemath{\bmx}{\mathbf{x}}
\safemath{\bmy}{\mathbf{y}}
\safemath{\bmz}{\mathbf{z}}
\safemath{\bmzero}{\mathbf{0}}
\safemath{\bmone}{\mathbf{1}}

\bmdefine{\biad}{a}
\bmdefine{\bibd}{b}
\bmdefine{\bicd}{c}
\bmdefine{\bidd}{d}
\bmdefine{\bied}{e}
\bmdefine{\bifd}{f}
\bmdefine{\bigd}{g}
\bmdefine{\bihd}{h}
\bmdefine{\biid}{i}
\bmdefine{\bijd}{j}
\bmdefine{\bikd}{k}
\bmdefine{\bild}{l}
\bmdefine{\bimd}{m}
\bmdefine{\bind}{n}
\bmdefine{\biod}{o}
\bmdefine{\bipd}{p}
\bmdefine{\biqd}{q}
\bmdefine{\bird}{r}
\bmdefine{\bisd}{s}
\bmdefine{\bitd}{t}
\bmdefine{\biud}{u}
\bmdefine{\bivd}{v}
\bmdefine{\biwd}{w}
\bmdefine{\bixd}{x}
\bmdefine{\biyd}{y}
\bmdefine{\bizd}{z}

\bmdefine{\bixid}{\xi}
\bmdefine{\bilambdad}{\lambda}
\bmdefine{\bimud}{\mu}
\bmdefine{\bithetad}{\theta}
\bmdefine{\biphid}{\phi}
\bmdefine{\bideltad}{\delta}

\safemath{\bmia}{\biad}
\safemath{\bmib}{\bibd}
\safemath{\bmic}{\bicd}
\safemath{\bmid}{\bidd}
\safemath{\bmie}{\bied}
\safemath{\bmif}{\bifd}
\safemath{\bmig}{\bigd}
\safemath{\bmih}{\bihd}
\safemath{\bmii}{\biid}
\safemath{\bmij}{\bijd}
\safemath{\bmik}{\bikd}
\safemath{\bmil}{\bild}
\safemath{\bmim}{\bimd}
\safemath{\bmin}{\bind}
\safemath{\bmio}{\biod}
\safemath{\bmip}{\bipd}
\safemath{\bmiq}{\biqd}
\safemath{\bmir}{\bird}
\safemath{\bmis}{\bisd}
\safemath{\bmit}{\bitd}
\safemath{\bmiu}{\biud}
\safemath{\bmiv}{\bivd}
\safemath{\bmiw}{\biwd}
\safemath{\bmix}{\bixd}
\safemath{\bmiy}{\biyd}
\safemath{\bmiz}{\bizd}

\safemath{\bmxi}{\bixid}
\safemath{\bmlambda}{\bilambdad}
\safemath{\bmmu}{\bimud}
\safemath{\bmtheta}{\bithetad}
\safemath{\bmphi}{\biphid}
\safemath{\bmdelta}{\bideltad}

\safemath{\bA}{\mathbf{A}}
\safemath{\bB}{\mathbf{B}}
\safemath{\bC}{\mathbf{C}}
\safemath{\bD}{\mathbf{D}}
\safemath{\bE}{\mathbf{E}}
\safemath{\bF}{\mathbf{F}}
\safemath{\bG}{\mathbf{G}}
\safemath{\bH}{\mathbf{H}}
\safemath{\bI}{\mathbf{I}}
\safemath{\bJ}{\mathbf{J}}
\safemath{\bK}{\mathbf{K}}
\safemath{\bL}{\mathbf{L}}
\safemath{\bM}{\mathbf{M}}
\safemath{\bN}{\mathbf{N}}
\safemath{\bO}{\mathbf{O}}
\safemath{\bP}{\mathbf{P}}
\safemath{\bQ}{\mathbf{Q}}
\safemath{\bR}{\mathbf{R}}
\safemath{\bS}{\mathbf{S}}
\safemath{\bT}{\mathbf{T}}
\safemath{\bU}{\mathbf{U}}
\safemath{\bV}{\mathbf{V}}
\safemath{\bW}{\mathbf{W}}
\safemath{\bX}{\mathbf{X}}
\safemath{\bY}{\mathbf{Y}}
\safemath{\bZ}{\mathbf{Z}}

\safemath{\bZero}{\mathbf{0}}
\safemath{\bOne}{\mathbf{1}}
\safemath{\bDelta}{\mathbf{\Delta}}
\safemath{\bLambda}{\mathbf{\UpLambda}}
\safemath{\bPhi}{\mathbf{\Upphi}}
\safemath{\bSigma}{\mathbf{\Upsigma}}
\safemath{\bOmega}{\mathbf{\Upomega}}
\safemath{\bTheta}{\mathbf{\Uptheta}}

\bmdefine{\biAd}{A}
\bmdefine{\biBd}{B}
\bmdefine{\biCd}{C}
\bmdefine{\biDd}{D}
\bmdefine{\biEd}{E}
\bmdefine{\biFd}{F}
\bmdefine{\biGd}{G}
\bmdefine{\biHd}{H}
\bmdefine{\biId}{I}
\bmdefine{\biJd}{J}
\bmdefine{\biKd}{K}
\bmdefine{\biLd}{L}
\bmdefine{\biMd}{M}
\bmdefine{\biOd}{N}
\bmdefine{\biPd}{O}
\bmdefine{\biQd}{P}
\bmdefine{\biRd}{R}
\bmdefine{\biSd}{S}
\bmdefine{\biTd}{T}
\bmdefine{\biUd}{U}
\bmdefine{\biVd}{V}
\bmdefine{\biWd}{W}
\bmdefine{\biXd}{X}
\bmdefine{\biYd}{Y}
\bmdefine{\biZd}{Z}

\bmdefine{\biDelta}{\Delta}
\bmdefine{\biLambda}{\Lambda}
\bmdefine{\biPhi}{\Phi}
\bmdefine{\biSigma}{\Sigma}
\bmdefine{\biOmega}{\Omega}
\bmdefine{\biTheta}{\Theta}

\safemath{\bimA}{\biAd}
\safemath{\bimB}{\biBd}
\safemath{\bimC}{\biCd}
\safemath{\bimD}{\biDd}
\safemath{\bimE}{\biEd}
\safemath{\bimF}{\biFd}
\safemath{\bimG}{\biGd}
\safemath{\bimH}{\biHd}
\safemath{\bimI}{\biId}
\safemath{\bimJ}{\biJd}
\safemath{\bimK}{\biKd}
\safemath{\bimL}{\biLd}
\safemath{\bimM}{\biMd}
\safemath{\bimN}{\biNd}
\safemath{\bimO}{\biOd}
\safemath{\bimP}{\biPd}
\safemath{\bimQ}{\biQd}
\safemath{\bimR}{\biRd}
\safemath{\bimS}{\biSd}
\safemath{\bimT}{\biTd}
\safemath{\bimU}{\biUd}
\safemath{\bimV}{\biVd}
\safemath{\bimW}{\biWd}
\safemath{\bimX}{\biXd}
\safemath{\bimY}{\biYd}
\safemath{\bimZ}{\biZd}

\safemath{\bimDelta}{\biDelta}
\safemath{\bimLambda}{\biLambda}
\safemath{\bimPhi}{\biPhi}
\safemath{\bimSigma}{\biSigma}
\safemath{\bimOmega}{\biOmega}
\safemath{\bimTheta}{\biTheta}

\safemath{\setA}{\mathcal{A}}
\safemath{\setB}{\mathcal{B}}
\safemath{\setC}{\mathcal{C}}
\safemath{\setD}{\mathcal{D}}
\safemath{\setE}{\mathcal{E}}
\safemath{\setF}{\mathcal{F}}
\safemath{\setG}{\mathcal{G}}
\safemath{\setH}{\mathcal{H}}
\safemath{\setI}{\mathcal{I}}
\safemath{\setJ}{\mathcal{J}}
\safemath{\setK}{\mathcal{K}}
\safemath{\setL}{\mathcal{L}}
\safemath{\setM}{\mathcal{M}}
\safemath{\setN}{\mathcal{N}}
\safemath{\setO}{\mathcal{O}}
\safemath{\setP}{\mathcal{P}}
\safemath{\setQ}{\mathcal{Q}}
\safemath{\setR}{\mathcal{R}}
\safemath{\setS}{\mathcal{S}}
\safemath{\setT}{\mathcal{T}}
\safemath{\setU}{\mathcal{U}}
\safemath{\setV}{\mathcal{V}}
\safemath{\setW}{\mathcal{W}}
\safemath{\setX}{\mathcal{X}}
\safemath{\setY}{\mathcal{Y}}
\safemath{\setZ}{\mathcal{Z}}
\safemath{\emptySet}{\varnothing}

\safemath{\colA}{\mathscr{A}}
\safemath{\colB}{\mathscr{B}}
\safemath{\colC}{\mathscr{C}}
\safemath{\colD}{\mathscr{D}}
\safemath{\colE}{\mathscr{E}}
\safemath{\colF}{\mathscr{F}}
\safemath{\colG}{\mathscr{G}}
\safemath{\colH}{\mathscr{H}}
\safemath{\colI}{\mathscr{I}}
\safemath{\colJ}{\mathscr{J}}
\safemath{\colK}{\mathscr{K}}
\safemath{\colL}{\mathscr{L}}
\safemath{\colM}{\mathscr{M}}
\safemath{\colN}{\mathscr{N}}
\safemath{\colO}{\mathscr{O}}
\safemath{\colP}{\mathscr{P}}
\safemath{\colQ}{\mathscr{Q}}
\safemath{\colR}{\mathscr{R}}
\safemath{\colS}{\mathscr{S}}
\safemath{\colT}{\mathscr{T}}
\safemath{\colU}{\mathscr{U}}
\safemath{\colV}{\mathscr{V}}
\safemath{\colW}{\mathscr{W}}
\safemath{\colX}{\mathscr{X}}
\safemath{\colY}{\mathscr{Y}}
\safemath{\colZ}{\mathscr{Z}}

\safemath{\opA}{\mathbb{A}}
\safemath{\opB}{\mathbb{B}}
\safemath{\opC}{\mathbb{C}}
\safemath{\opD}{\mathbb{D}}
\safemath{\opE}{\mathbb{E}}
\safemath{\opF}{\mathbb{F}}
\safemath{\opG}{\mathbb{G}}
\safemath{\opH}{\mathbb{H}}
\safemath{\opI}{\mathbb{I}}
\safemath{\opJ}{\mathbb{J}}
\safemath{\opK}{\mathbb{K}}
\safemath{\opL}{\mathbb{L}}
\safemath{\opM}{\mathbb{M}}
\safemath{\opN}{\mathbb{N}}
\safemath{\opO}{\mathbb{O}}
\safemath{\opP}{\mathbb{P}}
\safemath{\opQ}{\mathbb{Q}}
\safemath{\opR}{\mathbb{R}}
\safemath{\opS}{\mathbb{S}}
\safemath{\opT}{\mathbb{T}}
\safemath{\opU}{\mathbb{U}}
\safemath{\opV}{\mathbb{V}}
\safemath{\opW}{\mathbb{W}}
\safemath{\opX}{\mathbb{X}}
\safemath{\opY}{\mathbb{Y}}
\safemath{\opZ}{\mathbb{Z}}
\safemath{\opZero}{\mathbb{O}}
\safemath{\identityop}{\opI}

\safemath{\veca}{\bma}
\safemath{\vecb}{\bmb}
\safemath{\vecc}{\bmc}
\safemath{\vecd}{\bmd}
\safemath{\vece}{\bme}
\safemath{\vecf}{\bmf}
\safemath{\vecg}{\bmg}
\safemath{\vech}{\bmh}
\safemath{\veci}{\bmi}
\safemath{\vecj}{\bmj}
\safemath{\veck}{\bmk}
\safemath{\vecl}{\bml}
\safemath{\vecm}{\bmm}
\safemath{\vecn}{\bmn}
\safemath{\veco}{\bmo}
\safemath{\vecp}{\bmp}
\safemath{\vecq}{\bmq}
\safemath{\vecr}{\bmr}
\safemath{\vecs}{\bms}
\safemath{\vect}{\bmt}
\safemath{\vecu}{\bmu}
\safemath{\vecv}{\bmv}
\safemath{\vecw}{\bmw}
\safemath{\vecx}{\bmx}
\safemath{\vecy}{\bmy}
\safemath{\vecz}{\bmz}

\safemath{\veczero}{\bmzero}
\safemath{\vecone}{\bmone}
\safemath{\vecxi}{\bmxi}
\safemath{\veclambda}{\bmlambda}
\safemath{\vecmu}{\bmmu}
\safemath{\vectheta}{\bmtheta}
\safemath{\vecphi}{\bmphi}
\safemath{\vecdelta}{\bmdelta}

\safemath{\matA}{\bA}
\safemath{\matB}{\bB}
\safemath{\matC}{\bC}
\safemath{\matD}{\bD}
\safemath{\matE}{\bE}
\safemath{\matF}{\bF}
\safemath{\matG}{\bG}
\safemath{\matH}{\bH}
\safemath{\matI}{\bI}
\safemath{\matJ}{\bJ}
\safemath{\matK}{\bK}
\safemath{\matL}{\bL}
\safemath{\matM}{\bM}
\safemath{\matN}{\bN}
\safemath{\matO}{\bO}
\safemath{\matP}{\bP}
\safemath{\matQ}{\bQ}
\safemath{\matR}{\bR}
\safemath{\matS}{\bS}
\safemath{\matT}{\bT}
\safemath{\matU}{\bU}
\safemath{\matV}{\bV}
\safemath{\matW}{\bW}
\safemath{\matX}{\bX}
\safemath{\matY}{\bY}
\safemath{\matZ}{\bZ}
\safemath{\matzero}{\bmzero}

\safemath{\matDelta}{\bDelta}
\safemath{\matLambda}{\bLambda}
\safemath{\matPhi}{\bPhi}
\safemath{\matSigma}{\bSigma}
\safemath{\matOmega}{\bOmega}
\safemath{\matTheta}{\bTheta}

\safemath{\matidentity}{\matI}
\safemath{\matone}{\matO}

\safemath{\rnda}{A}
\safemath{\rndb}{B}
\safemath{\rndc}{C}
\safemath{\rndd}{D}
\safemath{\rnde}{E}
\safemath{\rndf}{F}
\safemath{\rndg}{G}
\safemath{\rndh}{H}
\safemath{\rndi}{I}
\safemath{\rndj}{J}
\safemath{\rndk}{K}
\safemath{\rndl}{L}
\safemath{\rndm}{M}
\safemath{\rndn}{N}
\safemath{\rndo}{O}
\safemath{\rndp}{P}
\safemath{\rndq}{Q}
\safemath{\rndr}{R}
\safemath{\rnds}{S}
\safemath{\rndt}{T}
\safemath{\rndu}{U}
\safemath{\rndv}{V}
\safemath{\rndw}{W}
\safemath{\rndx}{X}
\safemath{\rndy}{Y}
\safemath{\rndz}{Z}

\safemath{\rveca}{\bimA}
\safemath{\rvecb}{\bimB}
\safemath{\rvecc}{\bimC}
\safemath{\rvecd}{\bimD}
\safemath{\rvece}{\bimE}
\safemath{\rvecf}{\bimF}
\safemath{\rvecg}{\bimG}
\safemath{\rvech}{\bimH}
\safemath{\rveci}{\bimI}
\safemath{\rvecj}{\bimJ}
\safemath{\rveck}{\bimK}
\safemath{\rvecl}{\bimL}
\safemath{\rvecm}{\bimM}
\safemath{\rvecn}{\bimN}
\safemath{\rveco}{\bomO}
\safemath{\rvecp}{\bimP}
\safemath{\rvecq}{\bimQ}
\safemath{\rvecr}{\bimR}
\safemath{\rvecs}{\bimS}
\safemath{\rvect}{\bimT}
\safemath{\rvecu}{\bimU}
\safemath{\rvecv}{\bimV}
\safemath{\rvecw}{\bimW}
\safemath{\rvecx}{\bimX}
\safemath{\rvecy}{\bimY}
\safemath{\rvecz}{\bimZ}

\safemath{\rvecxi}{\bmxi}
\safemath{\rveclambda}{\bmlambda}
\safemath{\rvecmu}{\bmmu}
\safemath{\rvectheta}{\bmtheta}
\safemath{\rvecphi}{\bmphi}

\safemath{\rmatA}{\bimA}
\safemath{\rmatB}{\bimB}
\safemath{\rmatC}{\bimC}
\safemath{\rmatD}{\bimD}
\safemath{\rmatE}{\bimE}
\safemath{\rmatF}{\bimF}
\safemath{\rmatG}{\bimG}
\safemath{\rmatH}{\bimH}
\safemath{\rmatI}{\bimI}
\safemath{\rmatJ}{\bimJ}
\safemath{\rmatK}{\bimK}
\safemath{\rmatL}{\bimL}
\safemath{\rmatM}{\bimM}
\safemath{\rmatN}{\bimN}
\safemath{\rmatO}{\bimO}
\safemath{\rmatP}{\bimP}
\safemath{\rmatQ}{\bimQ}
\safemath{\rmatR}{\bimR}
\safemath{\rmatS}{\bimS}
\safemath{\rmatT}{\bimT}
\safemath{\rmatU}{\bimU}
\safemath{\rmatV}{\bimV}
\safemath{\rmatW}{\bimW}
\safemath{\rmatX}{\bimX}
\safemath{\rmatY}{\bimY}
\safemath{\rmatZ}{\bimZ}

\safemath{\rmatDelta}{\bimDelta}
\safemath{\rmatLambda}{\bimLambda}
\safemath{\rmatPhi}{\bimPhi}
\safemath{\rmatSigma}{\bimSigma}
\safemath{\rmatOmega}{\bimOmega}
\safemath{\rmatTheta}{\bimTheta}

\usepackage{amssymb}
\usepackage{amsfonts}
\usepackage{mathrsfs}
\usepackage{xspace}
\usepackage{bm}
\usepackage{fancyref}
\usepackage{textcomp}

\usepackage{multirow}
\usepackage{stmaryrd}

\newenvironment{textbmatrix}{	\setlength{\arraycolsep}{2.5pt}%
								\big[\begin{matrix}}{\end{matrix}\big]%
								\raisebox{0.08ex}{\vphantom{M}}}

\def\be{\begin{equation}}
\def\ee{\end{equation}}
\def\een{\nonumber \end{equation}}
\def\mat{\begin{bmatrix}}
\def\emat{\end{bmatrix}}
\def\btm{\begin{textbmatrix}}
\def\etm{\end{textbmatrix}}

\def\ba#1\ea{\begin{align}#1\end{align}}
\def\bas#1\eas{\begin{align*}#1\end{align*}}
\def\bs#1\es{\begin{split}#1\end{split}} 
\def\bg#1\eg{\begin{gather}#1\end{gather}}
\def\bml#1\eml{\begin{multline}#1\end{multline}}
\def\bi#1\ei{\begin{itemize}#1\end{itemize}}

\newcommand{\subtext}[1]{\text{\fontfamily{cmr}\fontshape{n}\fontseries{m}\selectfont{}#1}}
\newcommand{\lefto}{\mathopen{}\left}

\DeclareMathOperator{\tr}{tr}				
\DeclareMathOperator{\sign}{sgn}			
\DeclareMathOperator{\Exop}{\opE}			
\DeclareMathOperator{\Varop}{\mathrm{Var}}
\DeclareMathOperator{\Covop}{\mathrm{Cov}}
\DeclareMathOperator{\rot}{\nabla\times}		

\newcommand{\Ex}[2]{\ensuremath{\Exop_{#1}\lefto[#2\right]}} 	

\newcommand{\vecnorm}[1]{\lefto\lVert#1\right\rVert}		
\newcommand{\stupidnorm}[1]{\big\lVert#1\big\rVert}		
\newcommand{\opnorm}[1]{\lVert#1\rVert}		

\safemath{\dirac}{\delta}					
\safemath{\krond}{\dirac}					

\safemath{\upto}{\uparrow}
\safemath{\downto}{\downarrow}
\safemath{\iu}{j}							
\safemath{\ev}{\lambda}						
\safemath{\hilseqspace}{l^{2}}				
\newcommand{\banachfunspace}[1]{\setL^{#1}}	
\safemath{\hilfunspace}{\banachfunspace{2}}	

\safemath{\SNR}{\textsf{SNR}} 				
\safemath{\PAR}{\textsf{PAR}} 				
\safemath{\No}{N_0}							
\safemath{\Es}{E_s}							
\safemath{\Eb}{E_b}							
\safemath{\EbNo}{\frac{\Eb}{\No}}
\safemath{\EsNo}{\frac{\Es}{\No}}

\DeclareMathOperator{\CHop}{\ensuremath{\opH}} 
\safemath{\tvir}{\rndh_{\CHop}}				
\safemath{\tvtf}{\rndl_{\CHop}}				
\safemath{\spf}{\rnds_{\CHop}}				
\safemath{\bff}{H_{\CHop}}					

\safemath{\ircf}{r_{h}}						
\safemath{\tftvcf}{r_{s}}					
\safemath{\tfcf}{r_{l}}						
\safemath{\bfcf}{r_{H}}						

\safemath{\tcorr}{c_h}						
\safemath{\scf}{c_{s}}						
\safemath{\tfcorr}{c_{l}}					
\safemath{\fcorr}{c_{H}}						

\safemath{\mi}{I}							
\safemath{\capacity}{C}						

\safemath{\normal}{\mathcal{N}}			
\safemath{\jpg}{\mathcal{CN}}			
\safemath{\mchain}{\leftrightarrow}		
\newcommand{\AIC}{\ensuremath{\text{AIC}}} 

\safemath{\dB}{\,\mathrm{dB}}
\safemath{\dBm}{\,\mathrm{dBm}}
\safemath{\Hz}{\,\mathrm{Hz}}
\safemath{\kHz}{\,\mathrm{kHz}}
\safemath{\MHz}{\,\mathrm{MHz}}
\safemath{\GHz}{\,\mathrm{GHz}}
\safemath{\s}{\,\mathrm{s}}
\safemath{\ms}{\,\mathrm{ms}}
\safemath{\mus}{\,\mathrm{\text{\textmu}s}}
\safemath{\ns}{\,\mathrm{ns}}
\safemath{\ps}{\,\mathrm{ps}}
\safemath{\meter}{\,\mathrm{m}}
\safemath{\mm}{\,\mathrm{mm}}
\safemath{\cm}{\,\mathrm{cm}}
\safemath{\m}{\,\mathrm{m}}
\safemath{\W}{\,\mathrm{W}}
\safemath{\mW}{\, \mathrm{mW}}
\safemath{\J}{\,\mathrm{J}}
\safemath{\K}{\,\mathrm{K}}
\safemath{\bit}{\,\mathrm{bit}}
\safemath{\nat}{\,\mathrm{nat}}

\safemath{\define}{\triangleq}			

\safemath{\equivalent}{\sim}
\safemath{\distas}{\sim}					
\safemath{\sdiff}{\Delta}				

\safemath{\reals}{\mathbb{R}}
\safemath{\positivereals}{\reals_{+}}
\safemath{\integers}{\mathbb{Z}}
\safemath{\posint}{\integers_{+}}
\safemath{\naturals}{\mathbb{N}}
\safemath{\posnaturals}{\naturals_{+}}
\safemath{\complexset}{\mathbb{C}}
\safemath{\rationals}{\mathbb{Q}}

\newcommand*{\fancyrefapplabelprefix}{app}		
\newcommand*{\fancyrefthmlabelprefix}{thm}		
\newcommand*{\fancyreflemlabelprefix}{lem}		
\newcommand*{\fancyrefcorlabelprefix}{cor}		
\newcommand*{\fancyrefdeflabelprefix}{def}		
\newcommand*{\fancyrefproplabelprefix}{prop}	
\newcommand*{\fancyrefobslabelprefix}{obs}		
\newcommand*{\fancyrefalglabelprefix}{alg}		
\newcommand*{\fancyrefasmlabelprefix}{asm}	    
\newcommand*{\fancyreftbllabelprefix}{tab}	 

\frefformat{vario}{\fancyrefseclabelprefix}{Section~#1}
\frefformat{vario}{\fancyrefthmlabelprefix}{Theorem~#1}
\frefformat{vario}{\fancyreflemlabelprefix}{Lemma~#1}
\frefformat{vario}{\fancyrefcorlabelprefix}{Corollary~#1}
\frefformat{vario}{\fancyrefdeflabelprefix}{Definition~#1}
\frefformat{vario}{\fancyrefobslabelprefix}{Observation~#1}
\frefformat{vario}{\fancyrefasmlabelprefix}{Assumption~#1}
\frefformat{vario}{\fancyreffiglabelprefix}{Fig.~#1}
\frefformat{vario}{\fancyrefapplabelprefix}{Appendix~#1} 
\frefformat{vario}{\fancyrefproplabelprefix}{Proposition~#1}
\frefformat{vario}{\fancyrefalglabelprefix}{Algorithm~#1}
\frefformat{vario}{\fancyreftbllabelprefix}{Table~#1}
\frefformat{vario}{\fancyrefeqlabelprefix}{(#1)}

\newcommand{\marginnote}[1]{\marginpar{\framebox{\framebox{#1}}}}
\safemath{\dictab}{[\,\dicta\,\,\dictb\,]}

\safemath{\ysig}{\bmy}
\safemath{\ysighat}{\hat{\ysig}}
\safemath{\ysigdim}{M}
\safemath{\xsig}{\bmx}
\safemath{\xsigdim}{N}
\safemath{\nx}{n_x}
\safemath{\zsig}{\bmz}
\safemath{\zsigdim}{\ysigdim}
\safemath{\rsig}{\bmr}
\safemath{\Adict}{\bA}
\safemath{\Adicttilde}{\widetilde{\Adict}}
\safemath{\Adictdim}{\outputdim\times\xsigdim}
\safemath{\avec}{\bma}
\safemath{\avectilde}{\tilde{\avec}}
\safemath{\Bdict}{\bB}
\safemath{\Bdicttilde}{\widetilde{\Bdict}}
\safemath{\Cdict}{\bC}
\safemath{\cvec}{\bmc}
\safemath{\Ddict}{\bD}
\safemath{\Ddictdim}{\ysigdim\times\xsigdim}
\safemath{\dvec}{\bmd}
\safemath{\Ddicttilde}{\widetilde{\bD}}
\safemath{\Bonb}{\bB}
\safemath{\bvec}{\bmb}
\safemath{\Bonbdim}{\ysigdim\times\ysigdim}
\safemath{\noise}{\bmn}
\safemath{\noisedim}{\ysigim}
\safemath{\err}{\bme}
\safemath{\errdim}{\ysigdim}
\safemath{\errset}{\setE}
\safemath{\nerr}{n_e}
\safemath{\delop}{\bP_\errset}
\safemath{\delopc}{\bP_{{\errset}^c}}

\safemath{\cplxi}{\imath}
\safemath{\cplxj}{\jmath}

\safemath{\dict}{\matD}
\safemath{\inputdim}{N}		
\safemath{\outputdim}{M}		
\safemath{\sparsity}{S}	
\safemath{\inputdimA}{{N_a}}	
\safemath{\inputdimB}{{N_b}}	
\safemath{\elemA}{{n_a}}	
\safemath{\elemB}{{n_b}}	
\safemath{\resA}{\matR_a}	
\safemath{\resB}{\matR_b}	
\safemath{\subD}{\matS} 
\safemath{\subA}{\matS_a} 
\safemath{\subB}{\matS_b} 
\safemath{\dicta}{\matA} 	
\safemath{\dictb}{\matB} 	
\safemath{\hollowS}{H}
\safemath{\hollowA}{H_a}
\safemath{\hollowB}{H_b}
\safemath{\cross}{Z}
\safemath{\coh}{\mu_d}			
\safemath{\coha}{\mu_a}			
\safemath{\cohb}{\mu_b}			
\safemath{\mubs}{\nu}	
\safemath{\cohm}{\mu_m} 
\safemath{\dictset}{\setD}	
\safemath{\dictsetp}{\dictset(\coh,\coha,\cohb)}	
\safemath{\dictsetgen}{\dictset_\text{gen}}
\safemath{\dictsetgenp}{\dictsetgen(\coh)}
\safemath{\dictsetonb}{\dictset_\text{onb}}
\safemath{\dictsetonbp}{\dictsetonb(\coh)}

\safemath{\leftside}{U}
\safemath{\rightsideA}{R_a}
\safemath{\rightsideB}{R_b}

\safemath{\indexS}{\setI_S} 

\safemath{\na}{n_a}			
\safemath{\nb}{n_b}			
\safemath{\coeffa}{p_i}	
\safemath{\coeffb}{q_j}	
\safemath{\seta}{\setP}		
\safemath{\setb}{\setQ}     
\safemath{\setw}{\setW}	
\safemath{\setz}{\setZ}	
\safemath{\cola}{\veca}		
\safemath{\colb}{\vecb}		
\safemath{\cold}{\vecd}		
\safemath{\inputvec}{\vecx} 	
\safemath{\error}{\vece}	
\safemath{\noiseout}{\vecz} 	
\safemath{\inputvecel}{x}
\safemath{\inputveca}{\vecx_a}
\safemath{\inputvecb}{\vecx_b}
\safemath{\outputvec}{\vecy}	
\safemath{\lambdamin}{\lambda_{\mathrm{min}}}

\safemath{\elltwo}{\ell_2}
\safemath{\ellone}{\ell_1}
\safemath{\ellzero}{\ell_0}
\safemath{\ellinf}{\ell_\infty}
\safemath{\ellinftilde}{\ell_{\widetilde\infty}}
\safemath{\licard}{Z(\coh,\coha,\cohb)}
\safemath{\xsol}{\hat{x}}
\safemath{\xbord}{x_b}		
\safemath{\xstat}{x_s}		
\safemath{\xstatLone}{\tilde{x}_s}
\safemath{\order}{\mathcal{O}} 
\safemath{\scales}{\Theta} 
\safemath{\ones}{\mathbf{1}} 
\safemath{\zeroes}{\mathbf{0}} 
\safemath{\thlone}{\kappa(\coh,\cohb)} 
\safemath{\constoneA}{\delta} 
\safemath{\constoneB}{\epsilon} 
\safemath{\nlarge}{L}				   
\safemath{\sumlarge}{S_\nlarge}
\safemath{\maxlarger}{P_\nlarge}	   
\safemath{\Pzero}{\textrm{P0}}	
\safemath{\Pone}{\textrm{P1}}
\safemath{\vecfir}{\vecw}			 
\safemath{\vecsec}{\vecz}
\safemath{\elvecfir}{w}              
\safemath{\elvecsec}{z}				 
\safemath{\nlargefir}{n}
\safemath{\normout}{\gamma}
\safemath{\auxfun}{h}
\safemath{\supp}{\textrm{supp}}

\safemath{\indexa}{\ell}
\safemath{\indexb}{r}
\safemath{\indexc}{i}
\safemath{\indexd}{j}

\safemath{\project}{P}

\IEEEoverridecommandlockouts
\allowdisplaybreaks

\graphicspath{{./}}

\begin{document}

%
\title{Massive MU-MIMO-OFDM Uplink with \\ Direct RF-Sampling and 1-Bit ADCs}
\author{\IEEEauthorblockN{Sven Jacobsson$^\text{1,2}$, Lise Aabel$^\text{1,2}$, Mikael Coldrey$^\text{1}$, Ibrahim Can Sezgin$^\text{2}$, \\ Christian Fager$^\text{2}$, Giuseppe Durisi$^\text{2}$,  and Christoph Studer$^\text{3}$} \\
\thanks{The work of SJ and GD was supported in part by the Swedish Foundation for Strategic Research under grant ID14-0022, and by the Swedish Governmental Agency for Innovation Systems (VINNOVA) within the competence center ChaseOn. 
The work of ICZ and CF was supported by the Swedish Research Council under Grant 2015-04000.
The work of CS was supported in part by Xilinx, Inc.~and by the US National Science Foundation under grants ECCS-1408006, CCF-1535897,  CCF-1652065, CNS-1717559, and ECCS-1824379.}
\IEEEauthorblockA{
\footnotesize $^\text{1}$\textit{Ericsson Research, Gothenburg, Sweden}; $^\text{2}$\textit{Chalmers University of Technology, Gothenburg, Sweden}; $^\text{3}$\textit{Cornell Tech, New York City, NY, USA}
}
}

\maketitle

\begin{abstract}
\fussy 
Advances in analog-to-digital converter (ADC) technology have opened up the possibility to directly digitize wideband radio frequency~(RF) signals, avoiding the need for analog down-conversion. In this work, we consider an orthogonal frequency-division multiplexing (OFDM)-based massive multi-user~(MU) multiple-input multiple-output (MIMO) uplink system that relies on direct RF-sampling at the base station and digitizes the received RF signals with 1-bit ADCs. Using Bussgang's theorem, we provide an analytical expression for the error-vector magnitude (EVM) achieved by digital down-conversion and zero-forcing combining. Our results demonstrate that direct RF-sampling 1-bit ADCs enables low EVM and supports high-order constellations in the  massive MU-MIMO-OFDM uplink. 
\end{abstract}

\section{Introduction} \label{sec:intro}

Massive multi-user (MU) multiple-input multiple-output (MIMO) will be a key technology in upcoming cellular communication systems~\cite{boccardi14a}. 
This technology  enables significant gains in spectral efficiency and energy~efficiency~\cite{marzetta16a, rusek14a, swindlehurst14a} by equipping the base-station (BS) with a large number (e.g., hundreds) of antenna elements and serving multiple  user equipments (UEs) simultaneously in the same frequency band.

To fully exploit the advantages of massive MU-MIMO, each antenna element at the BS needs to be equipped with a set of analog-to-digital converters (ADCs) and digital-to-analog converters (DACs) to enable digital beamforming. 
However, to keep power consumption and system costs within tolerable limits when scaling up the number of antenna elements in such all-digital beamforming architectures, low-resolution (e.g., 1-to-6 bits) ADCs and DACs should be used at the BS, which inevitably deteriorates the system performance. Quite surprisingly, it has been shown that massive MU-MIMO is---up to some extent---robust against the imperfections caused by low-resolution ADCs and DACs at the BS~\cite{choi15a, wen15b, studer16a, liang16a, jeon18b, mollen16c, jacobsson18d, saxena16b, jacobsson17e, jacobsson17a, jedda17b, jacobsson17f, nedelcu17a, shao18a}.

\subsection{Homodyne Transceiver}

Existing theoretical results in~\cite{choi15a, wen15b, studer16a, liang16a, jeon18b, mollen16c, jacobsson18d, saxena16b, jacobsson17e, jacobsson17a, jedda17b, jacobsson17f, nedelcu17a, shao18a} that analyze low-resolution data converters (ADCs and DACs) in massive MU-MIMO implicitly assume that homodyne transceivers, also known as direct-conversion transceivers, are used at each antenna element at the BS. Homodyne transceivers perform down-conversion of radio frequency~(RF) signals to baseband~(BB) and up-conversion of BB signals to~RF in the analog domain.
In the receiver part of a homodyne transceiver, after analog down-conversion, the received BB signal is converted from the analog domain into the digital domain (an operation that involves sampling and quantization) by a pair of ADCs at each antenna element (one for the in-phase component and one for the quadrature component). 
In the transmitter part of a homodyne transceiver, before analog up-conversion, the in-phase and quadrature components of the transmitted BB signal are generated by a pair of~DACs. 


%

\begin{figure*}
\centering
\subfloat[Massive MU-MIMO-OFDM uplink with direct RF-sampling receiver chains at the BS. The per-antenna received RF signal is directly converted into the digital domain by a direct RF-sampling 1-bit ADC. After digital down-conversion (DDC), OFDM demodulation, and channel estimation, the received signals over the $B$ BS antennas are combined using a linear filter to detect the transmitted information.]{\includegraphics[width = 0.9\textwidth]{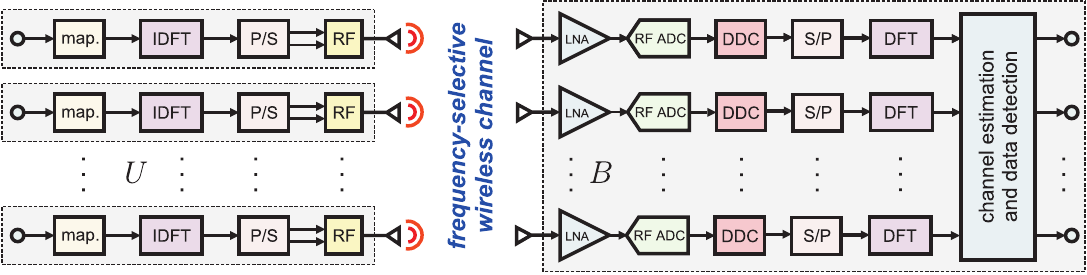}\label{fig:rf_sampling}}
\\
\subfloat[Massive MU-MIMO-OFDM uplink with homodyne receiver chains at the BS. At the BS, the per-antenna received RF signal is down-converted to BB by mixing it with an LO signal and converted into the digital domain by a pair of BB-sampling 1-bit ADCs. After OFDM demodulation and channel estimation, the received signals over the $B$ BS antennas are combined using a linear filter to detect the transmitted information.]{\includegraphics[width = 0.9\textwidth]{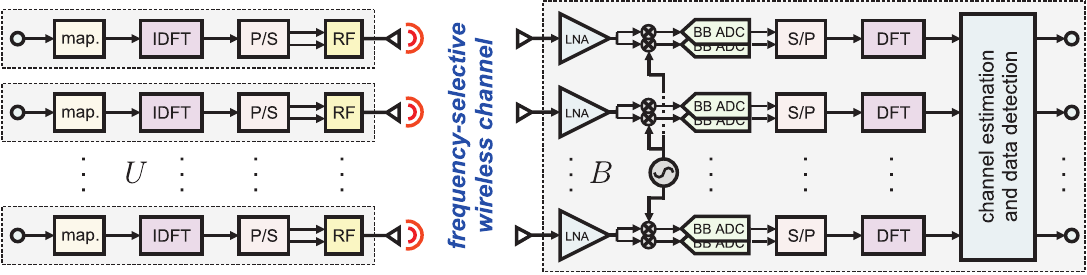}\label{fig:direct_conversion}}
\caption{Massive MU-MIMO-OFDM uplink system (excluding filters) where $U$ single-antenna UEs with ideal hardware independently perform symbol mapping and OFDM modulation. At the BS we consider (a) direct RF-sampling receiver chains and (b) homodyne receiver chains.}	
\label{fig:sys}
\end{figure*}

\subsection{Direct RF-Sampling Transceiver}

The sampling rates and energy efficiency of data converters are steadily improving every year~\cite{murmann08a, murmanna}, which opens up the possibility to design radio transceivers that perform analog-to-digital and digital-to-analog conversion directly in the RF domain, thus removing the need for analog up- and down-conversion. 
In such \emph{direct RF-sampling} transceivers, much of conventional RF circuitry, including local oscillators (LOs), filters and mixers, is replaced by digital signal processing (DSP), which enables simpler hardware designs, greater flexibility, and reduced system costs~\cite{neu15a, jayamohan15a, national-instruments19a, xilinx19a}.
%

In a direct RF-sampling transceiver, the sampling rate of the ADCs and DACs is typically of the order of several GS/s as it must be able to sample the frequency band of interest. Since the power consumption of data converters increases super-linearly with the sampling rate (for high-speed converters) and exponentially with the resolution~\cite{murmanna}, the design of low-resolution solutions for direct RF-sampling digital-beamforming systems is of paramount importance.
%
%
Hence, 1-bit direct RF transmitters have been proposed in~\cite{maehata13a,sezgin19a}.
In both these works, the 1-bit transmit waveform is generated by a band-pass $\Delta\Sigma$ modulator, which is implemented in a field-programmable gate array (FPGA), and fed to the antenna element over an optical interface.
A 1-bit direct RF-sampling single-antenna receiver has been presented in~\cite{prata16a}. In order to meet the error-vector magnitude (EVM) requirements set by the long-term evolution (LTE) and new radio (NR) standards~\cite{3gpp19b, 3gpp19a}, the received signal in~\cite{prata16a} is, prior to 1-bit quantization, dithered by a triangular waveform, which is subsequently subtracted from the quantized signal in the digital domain.
The results in \cite{maehata13a, sezgin19a, prata16a} demonstrate the feasibility of  direct RF-sampling transceivers with 1-bit ADCs and DACs. 
A theoretical analysis of the EVM achievable in such systems is, however,~missing.

\subsection{Contributions}

In this work, we consider a massive MU-MIMO uplink system in which the BS is equipped with direct RF-sampling 1-bit ADCs.
We focus on the scenario in which orthogonal frequency-division multiplexing (OFDM) is used to communicate over frequency-selective channels.
Using Bussgang's theorem~\cite{bussgang52a}, we derive an analytical expression for the EVM achieved by digital down-conversion (DDC) and zero-forcing~(ZF) combing.
We demonstrate that low values of EVM are attainable and high-order constellations are supported in the 1-bit direct RF-sampling massive MU-MIMO-OFDM uplink provided that the signal-to-noise ratio (SNR) is not too low and not too~high. 
We furthermore show that high-order constellations can be supported in the high-SNR regime by using a simple nonsubtractive dithering scheme. 

\subsection{Notation}

%
%
%
The $M \times 1$ all-zeros vector and the $M \times M$ identity~matrix are denoted by $\veczero_{M}$ and $\matI_M$, respectively.
%
%
The signum function $\sign(\cdot)$ is applied entry-wise to vectors and defined as $\sign(a)=1$ if $a \ge 0$ and $\sign(a)=-1$ if $a<0$.
The real part of a complex-valued vector $\veca$ is $\Re\{\veca\}$. 
The $\ell_2$-norm of $\veca$ is $\opnorm{\veca}_2$. 
The trace and main diagonal of a matrix $\matA$ is $\tr(\matA)$ and $\text{diag}(\matA)$, respectively.  
The pseudo-inverse of a tall matrix $\matA$ is $\matA^\dagger = (\matA^H\matA)^{-1}\matA^H$.
%
%
%
The real-valued zero-mean Gaussian distribution with covariance $\matR \in \opR^{M \times M}$ is $\normal(\veczero_M, \matR)$. The complex-valued circularly symmetric Gaussian distribution with covariance $\matC \in \opC^{M \times M}$ is $\jpg(\matzero_{M}, \matC)$.
The expected value of a random vector $\vecx$~is~$\Ex{}{\vecx}$. 
%
%

\section{Massive MU-MIMO-OFDM Uplink with \\ Direct RF-Sampling 1-Bit ADCs}

We consider a massive MU-MIMO-OFDM uplink system as depicted in~\fref{fig:rf_sampling}. Here, $U$ single-antenna UEs (with ideal hardware) transmit in the same time-frequency resource to a $B$-antenna BS that is equipped with direct RF-sampling 1-bit ADCs.
As a reference, \fref{fig:direct_conversion} depicts a massive MU-MIMO-OFDM uplink system with homodyne receiver chains and BB-sampling 1-bit ADCs at the BS. For the direct RF-sampling receiver, the LO and the mixers in~\fref{fig:direct_conversion} are replaced by a DDC stage in~\fref{fig:rf_sampling}, essentially moving complexity from the analog to the digital domain. 
%

%
%
%

\subsection{Channel Input-Output Model}

In what follows, we assume that all filters and low-noise amplifiers (LNAs) at the BS are ideal. We furthermore assume perfect timing and frequency synchronization between the BS and the~UEs---it has been shown in~\cite{stein17a, zhu18b, jacobsson19a} that accurate timing and frequency synchronization can be achieved even with low-resolution quantization.
Under these assumptions, the $n$th sample of the discrete-time $\text{1-bit}$ quantized RF signal received over the $B$ BS antennas can be written as follows:
%
%
\begin{IEEEeqnarray}{rCl} \label{eq:rx_1bit_rf}
	\vecz^\text{RF}_n &=& \sign\lefto( \vecy^\text{RF}_n \right)\!.
\end{IEEEeqnarray}
Here, $\vecy_n^\text{RF} \in \opR^{B}$ is the $n$th sample of the discrete-time RF signal, before 1-bit quantization, which we model as
\begin{IEEEeqnarray}{rCl} \label{eq:inout_pb}
\vecy^\text{RF}_n &=&  \vecx^\text{RF}_n + \vecw^\text{RF}_n
\end{IEEEeqnarray}
where
\begin{IEEEeqnarray}{rCl} \label{eq:pb_noiseless}
\vecx^\text{RF}_n 
&=& \sqrt{2}\,\Re\lefto\{ \vecx^\text{BB}_n  e^{j2\pi (f_\text{c}/f_\text{s}) n} \right\}\!.
\end{IEEEeqnarray}
Here, $f_\text{c}$ denotes the carrier frequency, $f_\text{s}$ denotes the sampling rate, and $\vecw^\text{RF}_n \distas \normal(\veczero_{B}, \frac{N_0}{2}\matI_B)$ is the additive white Gaussian noise (AWGN). The complex envelope $\vecx^\text{BB}_n \in \opC^B$ of $\vecx^\text{RF}_n \in \opR^B$ in~\eqref{eq:pb_noiseless} is given by
\begin{IEEEeqnarray}{rCl} \label{eq:xBB}
	\vecx^\text{BB}_n &=& \sum_{\ell = 0}^{L-1} \matH_\ell\vecs_{n-\ell}
\end{IEEEeqnarray}
where $\matH_\ell \in \opC^{B \times U}$, $\ell = 0,1,\dots,L-1$, is the $\ell$th tap of the frequency-selective channel connecting the $U$ UEs to the $B$-antenna BS.
Without loss of generality, we focus on the transmission of a single OFDM symbol for which the discrete-time transmit symbols $\vecs_n \in \opC^U$ in~\eqref{eq:xBB} are obtained through an inverse discrete Fourier transform (IDFT) as follows:  
\begin{IEEEeqnarray}{rCl} \label{eq:idft}
\vecs_n &=& \frac{1}{\sqrt{N}} \sum_{k \in \setS} \hat\vecs_k e^{j2\pi(k/N)n}, \quad n = 0,1,\dots,N-1. \IEEEeqnarraynumspace
\end{IEEEeqnarray}
Here,  $N \ge L$ is the total number of BS-side samples per OFDM symbol and $\setS \subset \{ 0, 1, \dots, N-1 \}$ is the set of occupied subcarriers. 
The vector $\hat\vecs_k \in \opC^U$ contains the transmitted frequency-domain symbols from the $U$ UEs. It holds that $\Ex{}{\hat\vecs_k\hat\vecs^H_k} = E_s \matI_U$ for $k \in \setS$ and $\vecs_k = \veczero_U$ for $k \notin \setS$.
To avoid interference between adjacent OFDM symbols, a cyclic prefix (CP) of $L-1$ samples is prepended to $\{ \vecs_n \}$. 

We define the signal-to-noise ratio (SNR) as $\textit{SNR} = E_s/N_0$, the signal bandwidth as $\textit{BW} = (S/N)f_\text{s}$ and the oversampling rate (OSR) of the direct RF-sampling 1-bit ADCs as $\textit{OSR} = N/S$. In direct RF-sampling systems, the sampling rate typically exceeds by far the signal bandwidth, i.e., we have $f_\text{s} \gg \textit{BW}$ such that $\textit{OSR} \gg 1$.

\subsection{Linear Decomposition using Bussgang's Theorem}

Bussgang's theorem is a simple yet powerful tool to analyze the impact of hardware impairments (see, e.g., \cite{mollen16c, jacobsson18d, saxena16b, jacobsson17e, jacobsson17a}). This theorem states that the correlation of two Gaussian signals, after one of them has undergone nonlinear distortion, is proportional to the correlation computed before the nonlinear distortion~\cite{bussgang52a}.
We will use this result to analyze the performance of direct RF-sampling with 1-bit ADCs.
In what follows, we assume that $\vecy^\text{RF}_n$ is Gaussian distributed, which is an accurate approximation for (i) OFDM signals or (ii) low SNR scenarios (where AWGN dominates). 
%
%
With Bussgang's theorem, the 1-bit quantized RF~signal in \eqref{eq:rx_1bit_rf}~becomes 
\begin{IEEEeqnarray}{rCl} \label{eq:decomp}
\vecz^\text{RF}_n 
&=& \matG \vecy^\text{RF}_n + \vece^\text{RF}_n =  \matG \vecx^\text{RF}_n + \matG \vecw^\text{RF}_n + \vece^\text{RF}_n
\end{IEEEeqnarray}
where the non-Gaussian RF distortion term $\vece^\text{RF}_n \in \opC^B$ is uncorrelated with the input $\vecy^\text{RF}_n$.
Furthermore, the diagonal matrix $\matG \in \opR^{B \times B}$ is given~by~\cite{mezghani12b}
\begin{IEEEeqnarray}{rCl} \label{eq:buss_gain_init}
	\matG &=& \sqrt{\frac{2}{\pi}} \matD_{\vecy^\text{RF}}^{-1/2}
\end{IEEEeqnarray} 
where $\matD_{\vecy^\text{RF}} = \text{diag}\lefto( \matR_{\vecy^\text{RF}}[0]\right) \in \opR^{B \times B}$. Here, $\matR_{\vecy^\text{RF}}[m] = \Ex{}{ \vecy^\text{RF}_n  (\vecy^\text{RF}_{n-m})^H} \in \opR^{B \times B}$ is the autocovariance of $\{ \vecy_n^\text{RF} \}$. 
To arrive at a closed-form expression for $\matG$, we shall first derive a closed-form expression for $\matR_{\vecy^\text{RF}}[m]$. 
It follows from~\eqref{eq:inout_pb} that
\begin{IEEEeqnarray}{rCl} \label{eq:CyRF}
	\matR_{\vecy^\text{RF}}[m] &=& 
\begin{cases}
	\matR_{\vecx^\text{RF}}[m] + \frac{N_0}{2} \matI_B, & m = 0 \\
	\matR_{\vecx^\text{RF}}[m], & \text{otherwise}.
\end{cases}
\end{IEEEeqnarray}
Here, $\matR_{\vecx^\text{RF}}[m] = \Ex{}{ \vecx^\text{RF}_n  (\vecx^\text{RF}_{n-m})^H} \in \opR^{B \times B}$. 
Since $\vecx^\text{BB}_n$ is a circularly symmetric random variable it follows from~\eqref{eq:pb_noiseless} and~\eqref{eq:xBB} that
\begin{IEEEeqnarray}{rCl} \label{eq:CxRF}
	\matR_{\vecx^\text{RF}}[m] &=& \Re\lefto\{\frac{E_s}{N} \sum_{k \in \setS}\widehat\matH_k\widehat\matH^H_k e^{j2\pi(k/N + f_\text{c}/f_\text{s})m} \right\}	
\end{IEEEeqnarray}
%
where $\widehat\matH_k = \sum_{\ell = 0}^{L-1} \matH_\ell e^{-j2\pi(k/N)\ell} \in \opC^{B \times U}$ is the frequency-domain channel matrix associated with the $k$th subcarrier. 
By inserting \eqref{eq:CyRF} and \eqref{eq:CxRF} into \eqref{eq:buss_gain_init}, we find a closed-form expression for $\matG$ as follows:
\begin{IEEEeqnarray}{rCl} \label{eq:G_final}
	\matG &=& \sqrt{\frac{2}{\pi}} \lefto(\text{diag}\lefto( \frac{E_s}{N} \sum_{k \in \setS}\widehat\matH_k\widehat\matH^H_k\right) + \frac{N_0}{2}\matI_B\right)^{\!\!-1/2}. \IEEEeqnarraynumspace
\end{IEEEeqnarray} 

\subsection{Digital Down-Conversion and Linear Combing}


Recall that in direct RF-sampling systems, the sampling rate typically exceeds the signal bandwidth by a large margin. 
Therefore, the received RF signal passes through a DDC stage (in which it is digitally filtered and down-sampled) prior to processing in the DSP unit.
In this work, we assume that the DDC stage is ideal, such that the frequency content associated with the set $\setS$ of occupied subcarriers is preserved perfectly and the rest is filtered out completely. 
Hence, the number of samples per OFDM symbol can be reduced from $N$ to $S$ without sacrificing performance and the received frequency-domain BB signal can be written~as follows:
\begin{IEEEeqnarray}{rCl} \label{eq:after_ddc}
\hat\vecz^\text{BB}_k 
&=& \sqrt{\frac{2}{N}} \sum_{n = 0}^{N-1} \vecz^\text{RF}_n e^{-j2\pi (k/N + f_\text{c}/f_\text{s}) n}, \quad k \in \setS. \IEEEeqnarraynumspace
\end{IEEEeqnarray}
By inserting \eqref{eq:decomp} into~\eqref{eq:after_ddc}, it can be shown that
\begin{IEEEeqnarray}{rCl} \label{eq:zkBB}
\hat\vecz^\text{BB}_k 
&=& \matG\widehat\matH_k\hat\vecs_k + \matG\hat\vecw^\text{BB}_k + \hat\vece^\text{BB}_k, \quad k \in \setS
\end{IEEEeqnarray}
where
\begin{IEEEeqnarray}{rCl} \label{eq:ekBB}
\hat\vece^\text{BB}_k 
&=& \sqrt{\frac{2}{N}} \sum_{n = 0}^{N-1} \vece^\text{RF}_n e^{-j2\pi (k/N + f_\text{c}/f_\text{s}) n}, \quad k \in \setS \IEEEeqnarraynumspace
\end{IEEEeqnarray}
is the frequency-domain BB distortion on the $k$th subcarrier. Furthermore, $\hat\vecw^\text{BB}_k = \sqrt{{2}/{N}} \sum_{n = 0}^{N-1} \vecw^\text{RF}_n e^{-j2\pi (k/N + f_\text{c}/f_\text{s}) n} \distas \jpg(\veczero_B, N_0\matI_B)$ is the AWGN on the $k$th subcarrier.


By assuming perfect channel state information (CSI) at the BS, the ZF estimate $\hat\vecs_k^\text{est}$ of the transmitted symbols $\hat\vecs_k$ is obtained from $\hat\vecz_k^\text{BB}$ as follows:
\begin{IEEEeqnarray}{rCl} \label{eq:zf}
	\hat\vecs^\text{est}_k &=& \widehat\matA_k \hat\vecz^\text{BB}_k, \quad k \in \setS.
\end{IEEEeqnarray}
Here, $\widehat\matA_k = ( \matG \widehat\matH_k )^\dagger \in \opC^{U \times B}$. By inserting \eqref{eq:zkBB} into~\eqref{eq:zf}, we finally obtain
\begin{IEEEeqnarray}{rCl} \label{eq:symb_est}
\hat\vecs^\text{est}_k
&=& \hat\vecs_k + \widehat\matA_k\matG\hat\vecw^\text{BB}_k + \widehat\matA_k\hat\vece^\text{BB}_k, \quad k \in \setS.
\end{IEEEeqnarray}
Note that this expression provides a \emph{linear} relationship between the input symbols and the output of the ZF combiner, which we next exploit when computing the EVM.


\section{Error-Vector Magnitude}
We define the EVM after DDC and ZF combining, averaged over the~$S$ occupied subcarriers and the $U$ symbol streams  as
\begin{IEEEeqnarray}{rCl} \label{eq:evm}
\textit{EVM} &=&  \sqrt{\frac{\sum_{k \in \setS}\Ex{}{\vecnorm{\hat\vecs^\text{est}_k - \hat\vecs_k}_2^2}}{\sum_{k \in \setS}\Ex{}{\vecnorm{\hat\vecs_k}_2^2}}} \cdot 100\%.
\end{IEEEeqnarray}
The denominator  is given by
\begin{IEEEeqnarray}{rCl} \label{eq:denominator}
	\sum_{k \in \setS}\Ex{}{\vecnorm{\hat\vecs_k}_2^2} &=& E_s U S.
\end{IEEEeqnarray}
The expected value in the numerator  can be expanded using~\eqref{eq:symb_est}, which yields
\begin{IEEEeqnarray}{rCl} 
\Ex{}{\vecnorm{\hat\vecs^\text{est}_k - \hat\vecs_k}_2^2} 
&=& \Ex{}{\stupidnorm{\widehat\matA_k\matG\hat\vecw^\text{BB}_k + \widehat\matA_k\hat\vece^\text{BB}_k}_2^2} \\
&=& \tr\lefto( \widehat\matA_k\lefto( N_0 \matG \matG + \matC_{\hat\vece_k^\text{BB}}\right) \widehat\matA_k^H \right) \IEEEeqnarraynumspace \label{eq:numerator}
\end{IEEEeqnarray}
where $\matC_{\hat\vece_k^\text{BB}} = \Ex{}{\hat\vece^\text{BB}_k(\hat\vece^\text{BB}_k)^H} \in \opC^{B \times B}$. 
Hence, to compute the EVM, we need a closed-form expression for $\matC_{\hat\vece_k^\text{BB}}$. 
It follows from~\eqref{eq:ekBB} that
\begin{IEEEeqnarray}{rCl} \label{eq:CekBB_1}
\matC_{\hat\vece^\text{BB}_k} 
&=& 2\sum_{m=0}^{N-1} \matR_{\vece^\text{RF}}[m] e^{-j2\pi(k/N + f_\text{c}/f_\text{s})m}
\end{IEEEeqnarray}
where $\matR_{\vece^\text{RF}}[m] = \Ex{}{ \vece^\text{RF}_n  (\vece^\text{RF}_{n-m})^H} \in \opR^{B \times B}$.
Since the vectors $\vecy_n^\text{RF}$ and $\vece_n^\text{RF}$ are uncorrelated, it follows from \eqref{eq:decomp}~that
\begin{IEEEeqnarray}{rCl} \label{eq:cov_uncorr}
	\matR_{\vece^\text{RF}}[m] &=& \matR_{\vecz^\text{RF}}[m] - \matG \matR_{\vecy^\text{RF}}[m] \matG.
\end{IEEEeqnarray}
Using Van Vleck's arcsine law~\cite{van-vleck66a}, we can write the autocovariance of the 1-bit quantized RF signal as follows:
\begin{IEEEeqnarray}{rCl} \label{eq:arcine}
	\matR_{\vecz^\text{RF}}[m] &=&	\frac{2}{\pi} \sin^{-1}\lefto( \matD_{\vecy^\text{RF}}^{-1/2} \matR_{\vecy^\text{RF}}[m] \matD_{\vecy^\text{RF}}^{-1/2} \right) \!.
\end{IEEEeqnarray}
Now, using \eqref{eq:cov_uncorr} and \eqref{eq:arcine}, we can rewrite \eqref{eq:CekBB_1} as follows:
\begin{IEEEeqnarray}{rCl}
\matC_{\hat\vece^\text{BB}_k} 
&=& 2\sum_{m=0}^{N-1} \bigg( \frac{2}{\pi} \sin^{-1}\lefto( \matD_{\vecy^\text{RF}}^{-1/2} \matR_{\vecy^\text{RF}}[m] \matD_{\vecy^\text{RF}}^{-1/2} \right)  \nonumber\\
&&  - \matG \matR_{\vecy^\text{RF}}[m] \matG \! \bigg) e^{-j2\pi(k/N + f_\text{c}/f_\text{s})m} \\
&=& \frac{4}{\pi} \sum_{m=0}^{N-1} \bigg( \! \sin^{-1}\lefto( \matD_{\vecy^\text{RF}}^{-1/2} \matR_{\vecy^\text{RF}}[m] \matD_{\vecy^\text{RF}}^{-1/2} \right) \nonumber\\
&& - \matD_{\vecy^\text{RF}}^{-1/2} \matR_{\vecy^\text{RF}}[m] \matD_{\vecy^\text{RF}}^{-1/2} \bigg)  e^{-j2\pi(k/N + f_\text{c}/f_\text{s})m}. \IEEEeqnarraynumspace \label{eq:CekBB_final}
\end{IEEEeqnarray}
By inserting \eqref{eq:G_final}, \eqref{eq:denominator}, \eqref{eq:numerator}, and \eqref{eq:CekBB_final} into \eqref{eq:evm} we arrive at an analytical expression for the EVM after DDC and ZF combing that depends on the second-order statistics of the input to the direct RF-sampling 1-bit ADCs, which in turn depends on the realization of the wireless~channel. 


\begin{figure}
\centering
\subfloat[1-bit ADCs ($\textit{EVM} = 8.2\%$).]{\includegraphics[width=.45\columnwidth]{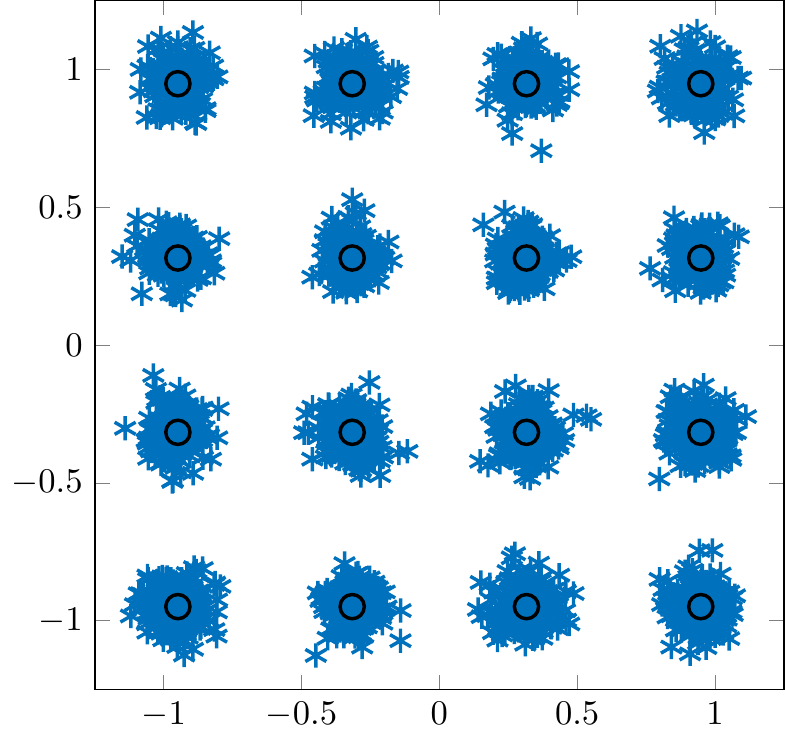}\label{fig:const_1bit}}	\qquad
\subfloat[Inf.~res.~($\textit{EVM} = 6.0\%$).]{\includegraphics[width=.45\columnwidth]{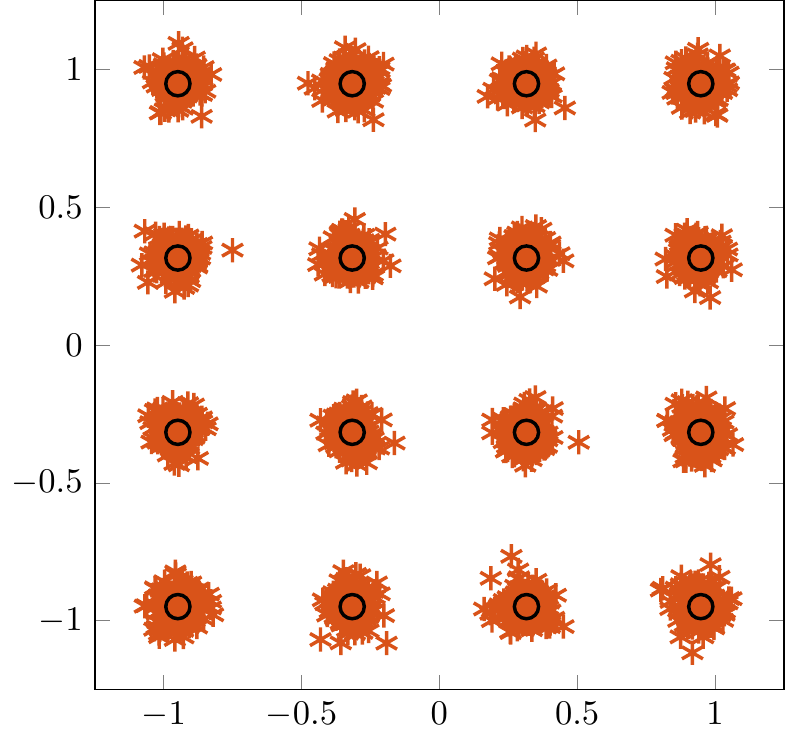}\label{fig:const_inf}}	
\caption{Received 16-QAM constellation after DDC and ZF combing. We consider a system with $B = 32$ antennas, $U=4$ UEs, $N=4096$ samples per OFDM symbol (excluding the CP), $S = 9$ occupied subcarriers, $L = 1000$ taps, $f_s = 10$\,GS/s, $f_c = 2.4$\,GHz, and $\textit{SNR} = 10$\,dB. The received 16-QAM symbols are clearly discernible despite the quantization error caused by the direct RF-sampling 1-bit ADCs.}
\label{fig:const}
\end{figure}

\section{Numerical Results} \label{sec:numerical}

We now show numerical results to confirm the validity of our analysis. Due to space constraints, we focus on a small set of parameters. 
Unless  stated otherwise, we consider a massive MU-MIMO-OFDM uplink system with $B = 32$ BS antennas and $U = 4$ UEs.
The carrier frequency is $f_\text{c} = 2.4$\,GHz and the sampling rate of the direct RF-sampling 1-bit ADCs is $f_\text{s} = 10$\,GS/s.
The number of samples per OFDM symbol is $N=4096$ and the number of occupied subcarriers is $S = 9$, which leads to $\textit{BW} \approx 22$\,MHz and $\textit{OSR} \approx 455$. The set of occupied subcarriers is $\setS = \{ 4093, 4094, 4095, 4096, 0, 1, 2, 3, 4\}$.
%
%
We average the EVM over $25$ random realizations of a frequency-selective Rayleigh fading channels with uniform power delay profile. The number of taps is $L=1000$, which corresponds to a delay spread of $L/f_\text{s} = 100$\,ns. For each channel realization, we send $25$ OFDM symbols with randomly generated 16-QAM symbols on the occupied~subcarriers.

As a proof-of-concept, we first show the received 16-QAM constellation after DDC and ZF combing  in~\fref{fig:const}.
We observe from \fref{fig:const_1bit} that, despite the quantization artifacts caused by direct RF-sampling with 1-bit ADCs, the constellation points in the 16-QAM constellation are clearly distinguishable. 
As a reference, the received constellation for the infinite-resolution (no quantization) case is shown in~\fref{fig:const_inf}.

\begin{figure}
\centering
\includegraphics[width=.9\columnwidth]{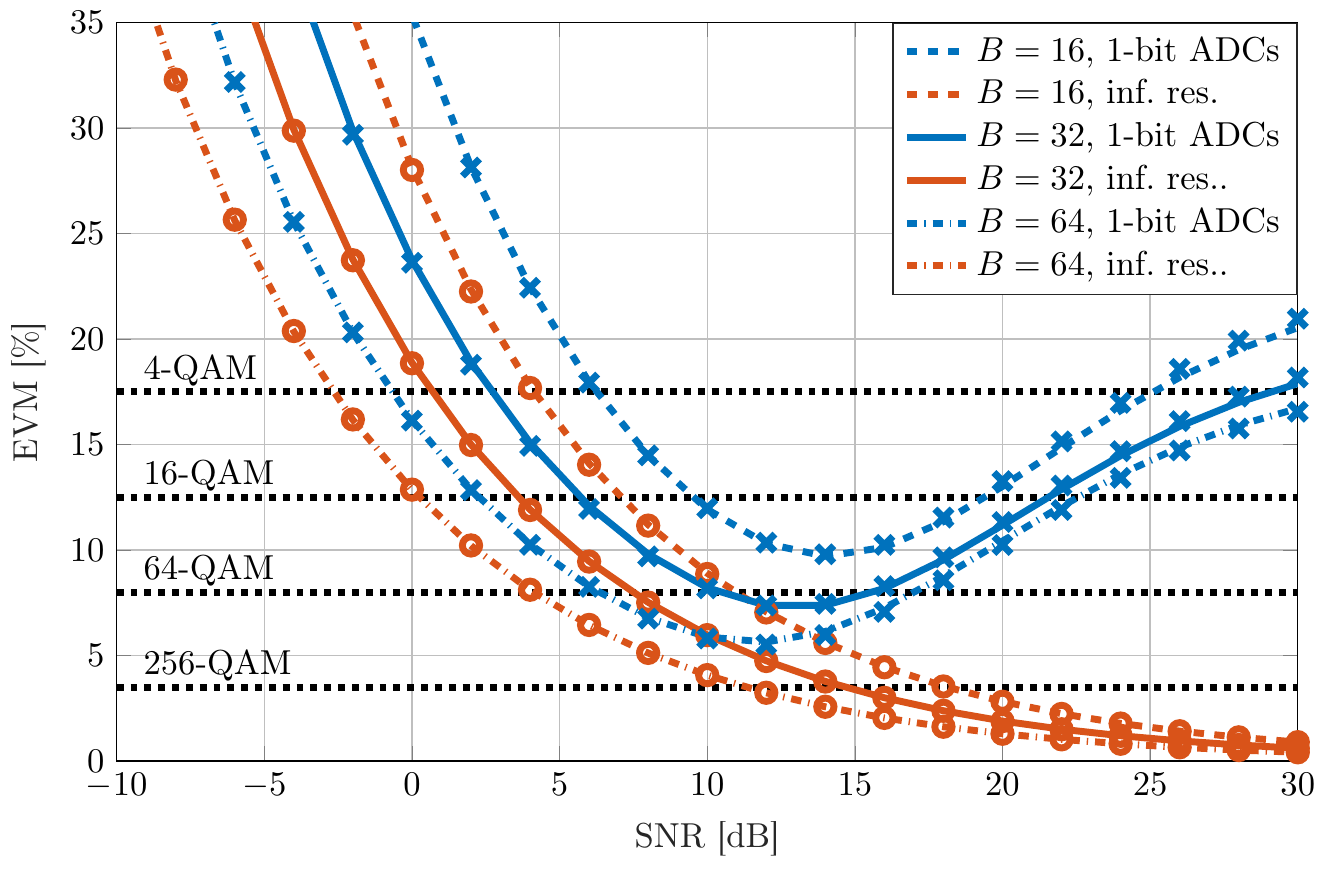}	
\caption{EVM after DDC and ZF combining. We consider a system with $B \in \{16, 32, 64\}$ antennas, $U=4$ UEs, $N=4096$ samples per OFDM symbol (excluding the CP), $S = 9$ occupied subcarriers, $L = 1000$ taps, $f_s = 10$\,GS/s, and $f_c = 2.4$\,GHz,. The markers correspond to simulation results and the lines to analytical results. 
Low values of EVM and high-order constellations are supported if the SNR is not too low and not too~high.
}
\label{fig:evm}
\end{figure}

\subsection{Error-Vector Magnitude}

In~\fref{fig:evm}, we show the EVM as a function of the SNR and the number of BS antennas.
We also provide the minimum EVM requirement for different modulation schemes according to the LTE and NR standards~\cite{3gpp19b, 3gpp19a}. 
We note that, in the 1-bit-ADC case, the EVM is not monotonously decreasing with the SNR.
This problem can be remedied by using nonsubtractive dithering, as we shall discuss in~\fref{sec:nonsub}.
We note that high-order constellations are supported if the SNR is not too low and not too high. For example, with $B = 32$ antennas, $64$-QAM is supported only for SNR values in the range 11\,dB-to-16\,dB. With $B = 64$ antennas, $64$-QAM is supported for SNR values in the wider range 7\,dB-to-17\,dB.
At low SNR, the analytical results matches perfectly the simulations. At high SNR, there is a slight discrepancy between the analytical results and the simulations stemming from the fact that the received signal is not perfectly Gaussian distributed, which is required by our analysis.

\subsection{Power Spectral Density}

\begin{figure*}
\centering
\subfloat[PSD of $\{\vecz^\text{RF}_n\}$ for the case $\textit{SNR} = 10$\,dB.]{\includegraphics[width=.3\textwidth]{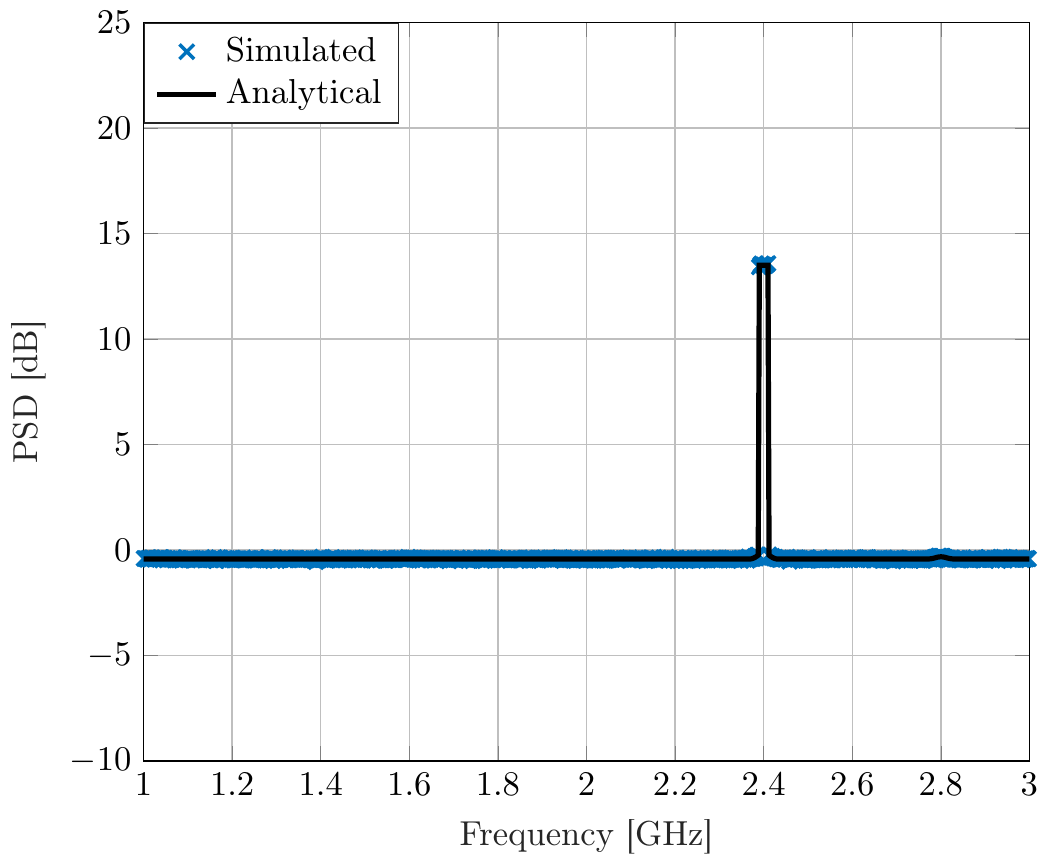}} \qquad
\subfloat[PSD of $\{\vecz^\text{RF}_n\}$ for the case $\textit{SNR} = 30$\,dB.]{\includegraphics[width=.3\textwidth]{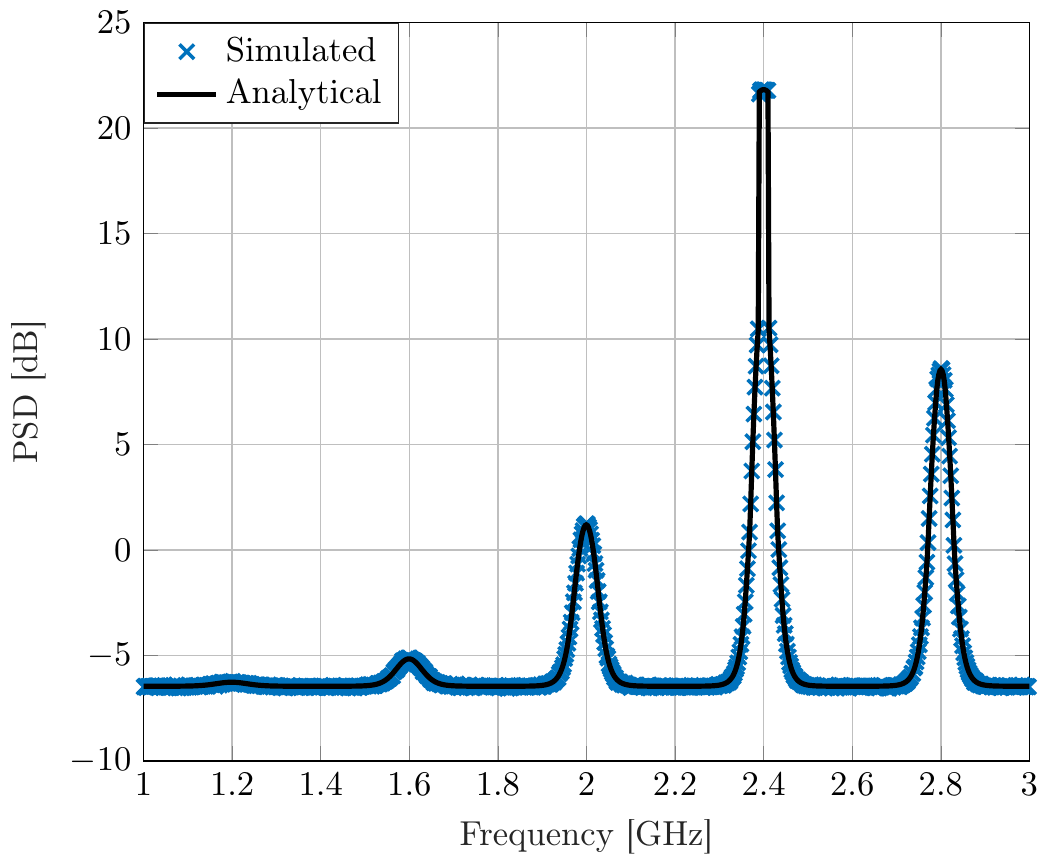}}	\qquad
\subfloat[PSD of $\{\vecz^\text{RF}_n\}$ for the case $\textit{SNR} = 50$\,dB.]{\includegraphics[width=.3\textwidth]{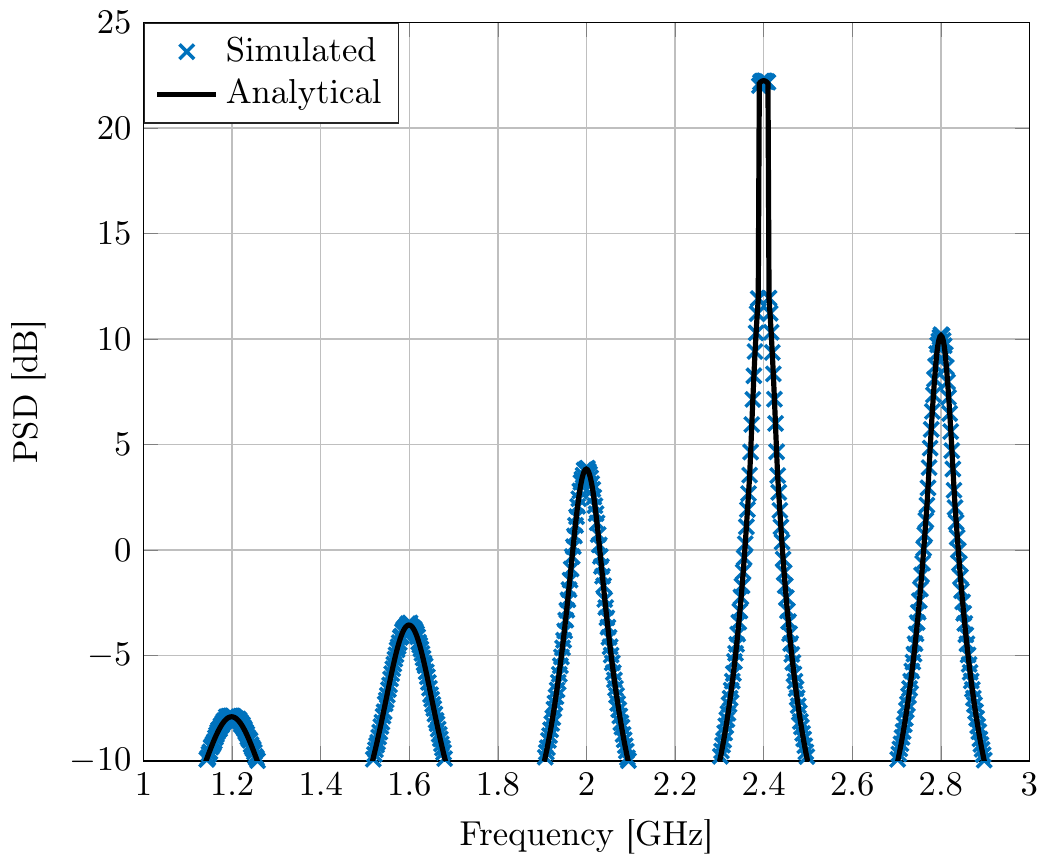}}	
\caption{PSD of the 1-bit quantized RF signal $\{\vecz^\text{RF}_n\}$ for different values of SNR. We consider a system with $B = 32$ antennas, $U=4$ UEs, $N=4096$ samples per OFDM symbol (excluding the CP), $S = 9$ occupied subcarriers, $L = 1000$ taps, $f_s = 10$\,GS/s, and $f_c = 2.4$\,GHz. The desired signal is clearly discernble at $f_c = 2.4$\,GHz. At high SNR, a larger portion of the distortion ends up in the same frequency band as the signal of interest.}
\label{fig:psd}
\end{figure*}

To gain insight into why the performance degrades with increasing SNR, we show the power spectral density (PSD) of the 1-bit quantized RF signal $\{ \vecz^\text{RF}_n\}$ for the frequency range $1$\,GHz-to-$3$\,GHz  in~\fref{fig:psd}. We obtain the analytical PSD by computing the discrete Fourier transform (DFT) of the autocovariance $\{\matR_{\vecz^\text{ZF}}[m]\}$. 
We start by noting that the signal of interest is clearly visible at $f_\text{c} = 2.4$\,GHz, even after 1-bit quantization.
As the SNR increases, more of the distortion ends up in the same frequency band as the signal of interest due to an increased temporal correlation in $\{ \vecy_n^\text{RF}\}$ and, hence, this distortion does not get filtered out in the DDC~stage.
%


\subsection{Nonsubtractive Dithering} \label{sec:nonsub}

%
From \eqref{eq:cov_uncorr} and \eqref{eq:arcine}, we see that temporal correlation in the quantizer input leads to temporal correlation in the distortion. Increasing the noise reduces the temporal correlation, which, in turn, means that less of the distortion ends up in the same frequency band as the signal of interest.
With dithering, i.e., intentionally adding noise to ``decorrelate'' the quantization error, the input to the direct RF-sampling 1-bit ADCs is
\begin{IEEEeqnarray}{rCl}
\vecy^\text{RF}_n &=&  \vecx^\text{RF}_n + \vecw^\text{RF}_n + \vecd^\text{RF}_n
\end{IEEEeqnarray}
where $\vecd_n^\text{RF} \in \opR^B$ is the dither signal.
In this work, we consider \emph{nonsubtractive} dithering (the dither signal is not subtracted in the digital domain and, hence, does not need to be known in the DSP unit).
Specifically, we consider uniform binary dither in which the entries of $\vecd_n^\text{RF}$ are drawn uniformly from the set $\{ \pm \sqrt{D_0/2}\}$ and Gaussian dither in which the entries of $\vecd_n^\text{RF}$ are drawn from a $\normal(0, D_0/2)$ distribution. 
We note that uniform binary dithering could be implemented using 1-bit DACs.
The autocovariance of $\{ \vecy_n^\text{RF}\}$ in \eqref{eq:CyRF} becomes
\begin{IEEEeqnarray}{rCl} \label{eq:CyRF_dither}
	\matR_{\vecy^\text{RF}}[m] &=& 
\begin{cases}
	\matR_{\vecx^\text{RF}}[m] + \frac{N_0 + D_0}{2} \matI_B, & m = 0 \\
	\matR_{\vecx^\text{RF}}[m], & \text{otherwise}.
\end{cases}
\end{IEEEeqnarray}
%
%
By replacing $N_0$ with $N_0 + D_0$ in our analysis, we find an~expression for the EVM with nonsubtractive Gaussian~dithering. 

\begin{figure}
\centering
\includegraphics[width=.9\columnwidth]{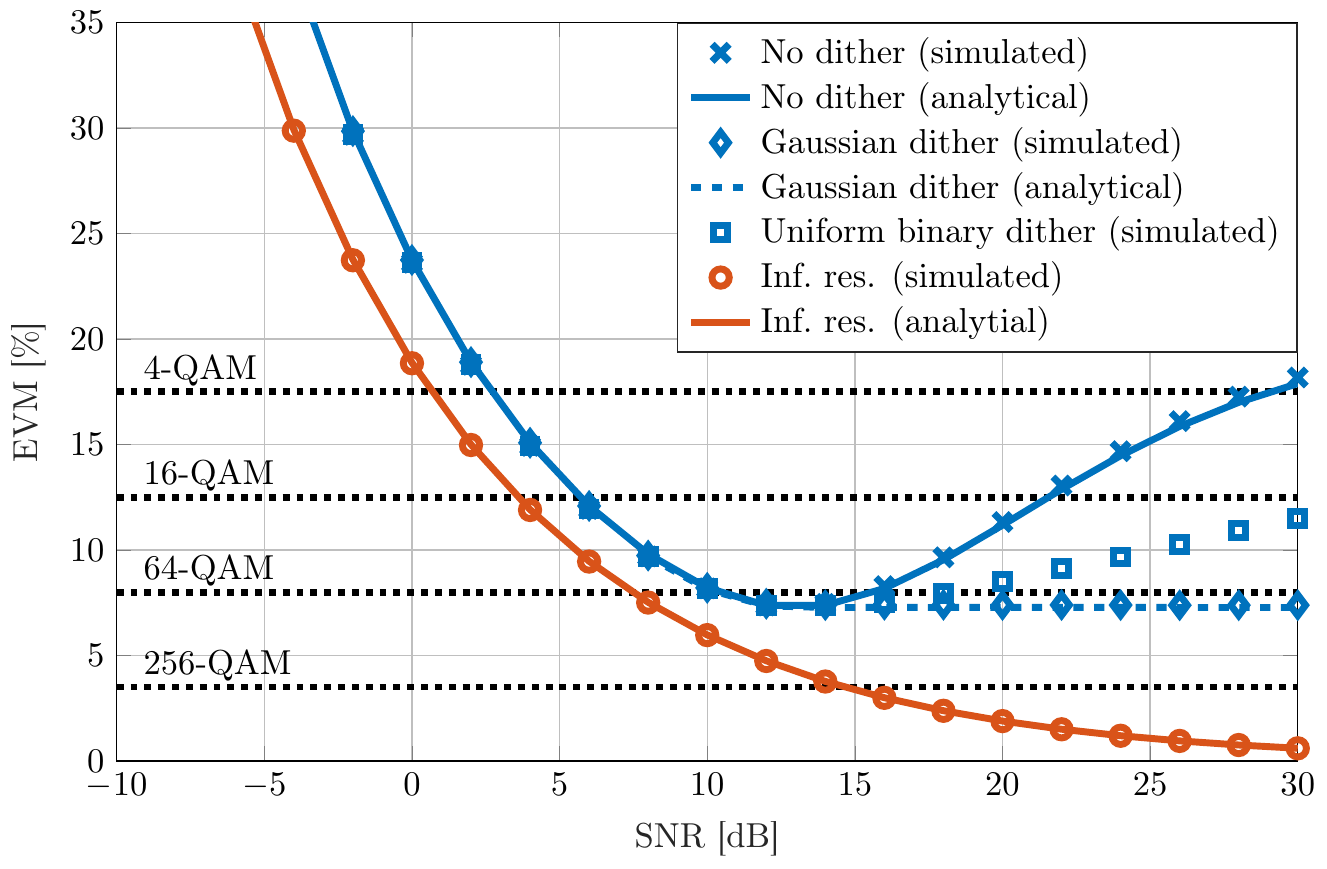}	
\caption{EVM with and without nonsubtractive dither after DDC and ZF combining. We consider a system with $B =32$ antennas, $U=4$ UEs, $N=4096$ samples per OFDM symbol (excluding the CP), $S = 9$ occupied subcarriers, $L = 1000$ taps, $f_s = 10$\,GS/s, and $f_c = 2.4$\,GHz. Higher-order constellations are supported at high SNR with nonsubtractive dithering.}
\label{fig:dither}
\end{figure}

In~\fref{fig:dither}, we show the EVM with and without nonsubtractive dithering. We have optimized $D_0$ for each SNR value. With Gaussian dithering, the EVM is nonincreasing in SNR; with uniform binary dithering, the EVM is lower than that of the nondithered case but still increases at high~SNR.


\section{Conclusions}

We have shown that low values of EVM and high-order constellations can be achieved in a massive MU-MIMO-OFDM uplink system in which the BS is equipped with direct RF-sampling 1-bit ADCs.
We have derived an analytical expression for the EVM after DDC and ZF combing by leveraging Bussgang's theorem and Van Vleck's arcsine law.

There exist many avenues for future work. 
Demonstrating of a cost-effective, real-world implementation of direct RF-sampling in multi-antenna transceivers is part of ongoing work.
A key concern for direct RF-sampling receivers is the sensitivity to blocking~\cite{neu15a}; a corresponding study is part of future work.
The high-SNR performance can likely be considerably improved by choosing a nonsubtractive or subtractive dither signal that is tailored to the problem at hand; designing such dithering schemes is left for future work.


\balance

\begin{spacing}{.96}
\bibliographystyle{IEEEtran}
\bibliography{IEEEabrv,confs-jrnls,publishers,svenbib}		

\begin{thebibliography}{10}
\providecommand{\url}[1]{#1}
\csname url@samestyle\endcsname
\providecommand{\newblock}{\relax}
\providecommand{\bibinfo}[2]{#2}
\providecommand{\BIBentrySTDinterwordspacing}{\spaceskip=0pt\relax}
\providecommand{\BIBentryALTinterwordstretchfactor}{4}
\providecommand{\BIBentryALTinterwordspacing}{\spaceskip=\fontdimen2\font plus
\BIBentryALTinterwordstretchfactor\fontdimen3\font minus
  \fontdimen4\font\relax}
\providecommand{\BIBforeignlanguage}[2]{{%
\expandafter\ifx\csname l@#1\endcsname\relax
\typeout{** WARNING: IEEEtran.bst: No hyphenation pattern has been}%
\typeout{** loaded for the language `#1'. Using the pattern for}%
\typeout{** the default language instead.}%
\else
\language=\csname l@#1\endcsname
\fi
#2}}
\providecommand{\BIBdecl}{\relax}
\BIBdecl

\bibitem{boccardi14a}
F.~Boccardi, R.~W. {Heath Jr.}, A.~Lozano, T.~L. Marzetta, and P.~Popovski,
  ``Five disruptive technology directions for {5G},'' \emph{{IEEE} Commun.
  Mag.}, vol.~52, no.~2, pp. 74--80, Feb. 2014.

\bibitem{marzetta16a}
T.~L. Marzetta, E.~G. Larsson, H.~Yang, and H.~Q. Ngo, \emph{Fundamentals of
  Massive {MIMO}}.\hskip 1em plus 0.5em minus 0.4em\relax Cambridge Univ.
  Press, 2016.

\bibitem{rusek14a}
F.~Rusek, D.~Persson, B.~Kiong, E.~G. Larsson, T.~L. Marzetta, O.~Edfors, and
  F.~Tufvesson, ``Scaling up {MIMO}: Opportunities and challenges with very
  large large arrays,'' \emph{{IEEE} Signal Process. Mag.}, vol.~30, no.~1, pp.
  40--60, Jan. 2013.

\bibitem{swindlehurst14a}
A.~L. Swindlehurst, E.~Ayanoglu, P.~Heydari, and F.~Capolino, ``Millimeter-wave
  massive {MIMO}: The next wireless revolution?'' \emph{{IEEE} Commun. Mag.},
  vol.~52, no.~9, pp. 56--62, Sep. 2014.

\bibitem{choi15a}
J.~Choi, J.~Mo, and R.~W. {Heath Jr.}, ``Near maximum-likelihood detector and
  channel estimator for uplink multiuser massive {MIMO} systems with one-bit
  {ADC}s,'' \emph{{IEEE} Trans. Commun.}, vol.~64, no.~5, pp. 2005--2018, May
  2016.

\bibitem{wen15b}
C.-K. Wen, C.-J. Wang, S.~Jin, K.-K. Wong, and P.~Ting, ``{B}ayes-optimal joint
  channel-and-data estimation for massive {MIMO} with low-precision {ADC}s,''
  \emph{{IEEE} Trans. Signal Process.}, vol.~64, no.~10, pp. 2541--2556, Jul.
  2015.

\bibitem{studer16a}
C.~Studer and G.~Durisi, ``Quantized massive {MU-MIMO-OFDM} uplink,''
  \emph{{IEEE} Trans. Commun.}, vol.~64, no.~6, pp. 2387--2399, Jun. 2016.

\bibitem{liang16a}
N.~Liang and W.~Zhang, ``Mixed-{ADC} massive {MIMO} uplink in
  frequency-selective channels,'' \emph{{IEEE} Trans. Commun.}, vol.~64,
  no.~11, pp. 4652--4666, Sep. 2016.

\bibitem{jeon18b}
Y.-S. Jeon, N.~Lee, S.-N. Hong, and R.~W. {Heath Jr.}, ``One-bit sphere
  decoding for uplink massive mimo systems with one-bit {ADCs},'' \emph{{IEEE}
  Trans. Wireless Commun.}, vol.~17, no.~7, pp. 4509--4521, Jul. 2018.

\bibitem{mollen16c}
C.~Moll{\'e}n, J.~Choi, E.~G. Larsson, and R.~W. {Heath Jr.}, ``Uplink
  performance of wideband massive {MIMO} with one-bit {ADCs},'' \emph{{IEEE}
  Trans. Wireless Commun.}, vol.~16, no.~1, pp. 87--100, Jan. 2017.

\bibitem{jacobsson18d}
S.~Jacobsson, U.~Gustavsson, G.~Durisi, and C.~Studer, ``Massive
  {MU}-{MIMO}-{OFDM} uplink with hardware impairments: Modeling and analysis,''
  in \emph{Proc. Asilomar Conf. Signals, Syst., Comput.}, Pacific Grove, CA,
  USA, Oct. 2018, pp. 1829--1835.

\bibitem{saxena16b}
A.~K. Saxena, I.~Fijalkow, and A.~L. Swindlehurst, ``Analysis of one-bit
  quantized precoding for the multiuser massive {MIMO} downlink,'' \emph{{IEEE}
  Trans. Signal Process.}, vol.~65, no.~17, pp. 4624--4634, Sep. 2017.

\bibitem{jacobsson17e}
S.~Jacobsson, G.~Durisi, M.~Coldrey, and C.~Studer, ``Linear precoding with
  low-resolution {DACs} for massive {MU}-{MIMO}-{OFDM} downlink,'' \emph{{IEEE}
  Trans. Wireless Commun.}, vol.~18, no.~3, pp. 1595--1609, Mar. 2019.

\bibitem{jacobsson17a}
S.~Jacobsson, M.~Coldrey, G.~Durisi, and C.~Studer, ``On out-of-band emissions
  of quantized precoding in massive {MU-MIMO-OFDM},'' in \emph{Proc. Asilomar
  Conf. Signals, Syst., Comput.}, Pacific Grove, CA, USA, Oct.--Nov. 2017, pp.
  21--26.

\bibitem{jedda17b}
H.~Jedda, A.~Mezghani, J.~A. Nossek, and A.~L. Swindlehurst, ``Massive {MIMO}
  downlink 1-bit precoding for frequency selective channels,'' in \emph{Int.
  Workshop Comput. Advances Multi-Sensor Adaptive Process. (CAMSAP)}, Curacao,
  Curacao, Dec. 2017.

\bibitem{jacobsson17f}
S.~Jacobsson, O.~Casta\~{n}eda, C.~Jeon, G.~Durisi, and C.~Studer, ``Nonlinear
  precoding for phase-quantized constant-envelope massive {MU}-{MIMO}-{OFDM},''
  in \emph{Proc. {IEEE} Int. Conf. Telecommunications (ICT)}, St.~Malo, France,
  Jun. 2018, pp. 367--372.

\bibitem{nedelcu17a}
A.~Nedelcu, F.~Steiner, M.~Staudacher, G.~Kramer, W.~Zirwas, R.~Sisava~Ganesan,
  P.~Baracca, and S.~Wesemann, ``Quantized precoding for multi-antenna downlink
  channels with {MAGIQ},'' in \emph{Int. ITG Workshop on Smart Antennas (WSA)},
  Bochum, Germany, Mar. 2017.

\bibitem{shao18a}
M.~Shao, Q.~Li, and W.-K. Ma, ``One-bit massive {MIMO} precoding via minimum
  symbol-error probability design,'' in \emph{Proc. IEEE Int. Conf. Acoust.,
  Speech, Signal Process. (ICASSP)}, Calgary, AB, Canada, Mar. 2018, pp.
  3579--3583.

\bibitem{murmann08a}
B.~Murmann, ``{A/D} converter trends: Power dissipation, scaling and digitally
  assisted architectures,'' in \emph{Proc. IEEE Custom Integrated Circuits
  Conf. (CICC)}, San Jose, CA, USA, Sep. 2008, pp. 105--112.

\bibitem{murmanna}
\BIBentryALTinterwordspacing
------, ``{ADC} performance survey 1997-2019.'' [Online]. Available:
  \url{http://web.stanford.edu/~murmann/adcsurvey.html}
\BIBentrySTDinterwordspacing

\bibitem{neu15a}
T.~Neu, ``Direct {RF} conversion: From vision to reality,'' 2015, {Texas
  Instruments}.

\bibitem{jayamohan15a}
U.~Jayamohan, ``Not your grandfather's {ADC}: {RF} sampling {ADCs} offer
  advantages in system design,'' 2015, {Analog Devices}.

\bibitem{national-instruments19a}
{National Instruments}, ``Advantages of direct {RF} sampling architectures,''
  White Paper, May 2019.

\bibitem{xilinx19a}
Xilinx, ``An adaptable direct {RF}-sampling solution,'' White Paper, Feb. 2019.

\bibitem{maehata13a}
T.~Maehata, K.~Totani, T.~Asaina, and H.~Tachibana, ``Development of 1-bit
  digital radio-frequency transmitter,'' \emph{SEI Technical Review}, no.~76,
  pp. 84--89, Apr. 2013.

\bibitem{sezgin19a}
I.~C. Sezgin, M.~Dahlgren, T.~Eriksson, M.~Coldrey, C.~Larsson, J.~Gustavsson,
  and C.~Fager, ``A low-complexity distributed-{MIMO} testbed based on
  high-speed sigma--delta-over-fiber,'' \emph{{IEEE} Trans. Microw. Theory
  Techn.}, 2019, to appear.

\bibitem{prata16a}
A.~Prata, A.~S.~R. Oliviera, and N.~B. Carvalho, ``All-digital flexible uplink
  remote radio head for {C}-{RAN},'' in \emph{Proc. {IEEE} {MTTS} Int. Microw.
  Symp. (IMS)}, San Fransisco, CA, USA, May 2016.

\bibitem{3gpp19b}
3GPP, ``{LTE}; evolved universal terrestrial radio access ({E-UTRA}); base
  station ({BS}) radio transmission and reception,'' May 2019, {TS} 36.104
  version 12.13.0 Rel.~12.

\bibitem{3gpp19a}
------, ``{5G}; {NR}; base station ({BS}) radio transmission and reception,''
  May 2019, {TS} 38.104 version 15.5.0 Rel.~15.

\bibitem{bussgang52a}
J.~J. Bussgang, ``Crosscorrelation functions of amplitude-distorted {Gaussian}
  signals,'' Res. Lab. Elec., Cambridge, MA, USA, Tech. Rep. 216, Mar. 1952.

\bibitem{stein17a}
M.~S. Stein, ``Performance analysis for time-of-arrival estimation with
  oversampled low-complexity 1-bit {A/D} conversion,'' in \emph{Proc. IEEE Int.
  Conf. Acoust., Speech, Signal Process. (ICASSP)}, New Orleans, LA, USA, Mar.
  2017, pp. 4491--4495.

\bibitem{zhu18b}
\BIBentryALTinterwordspacing
D.~Zhu, R.~Bendlin, S.~Akoum, A.~Ghosh, and R.~W. {Heath Jr.}, ``Directional
  frame timing synchronization in wideband millimeter-wave systems with
  low-resolution {ADCs},'' Sep. 2018. [Online]. Available:
  \url{https://arxiv.org/abs/1809.02890}
\BIBentrySTDinterwordspacing

\bibitem{jacobsson19a}
S.~Jacobsson, C.~Lindquist, G.~Durisi, T.~Eriksson, and C.~Studer, ``Timing and
  frequency synchronization for 1-bit massive {MU}-{MIMO}-{OFDM} downlink,'' in
  \emph{IEEE Int. Workshop Signal Process. Advances Wireless Commun. (SPAWC)},
  Cannes, France, Jul. 2019, to appear.

\bibitem{mezghani12b}
A.~Mezghani and J.~A. Nossek, ``Capacity lower bound of {MIMO} channels with
  output quantization and correlated noise,'' in \emph{IEEE Int. Symp. Inf.
  Theory (ISIT)}, Cambridge, MA, USA, Jul. 2012.

\bibitem{van-vleck66a}
J.~H. Van~Vleck and D.~Middleton, ``The spectrum of clipped noise,''
  \emph{Proc. {IEEE}}, vol.~54, no.~1, pp. 2--19, Jan. 1966.

\end{thebibliography}
\end{spacing}

\balance

\end{document}